

\magnification=1200
\baselineskip=20pt
\hfil{NHCU-HEP-94-28}
\centerline{\bf Classical Extended Conformal Algebras Associated with}
\centerline{\bf Constrained KP Hierarchy}
\vskip 1.5cm
\centerline{Wen-Jui Huang$^1$, J.C. Shaw$^2$ and H.C. Yen$^1$}
\vskip 1 cm
\centerline{$^1$Department of Physics, National Tsing Hua University}
\centerline{Hsinchu, Taiwan R.O.C.}
\centerline{$^2$Department of Applied Mathematics, National Chiao Tung
University}
\centerline{Hsinchu, Taiwan R.O.C.}
\vskip 1cm
\centerline{\bf Abstract}
\vskip 0.5cm

We examine the conformal preperty of the second Hamiltonian structure of
constrained KP hierarchy derived by Oevel and Strampp. We find that it
naturally gives a family of nonlocal extended conformal algebras. We give
two examples of such algebras and find that they are similar to
 Bilal's V algebra. By taking a gauge
transformation one can map the constrained KP hierarchy to Kupershmidt's
nonstardard Lax hierarchy. We consider the second Hamiltonian structure
in this representation. We show that after mapping the Lax operator to a
pure differential operator the second structure becomes the sum of the second
and the third Gelfand-Dickey brackets defined by this differential operator.
We show that this Hamiltonian structure defines the W-U(1)-Kac-Moody algebra
by working out its conformally covariant form.
\vskip .5cm

\noindent{PACS: 03.40.-t, 11.30.Pb}
\vfil\eject

\noindent{\bf I. Introduction}

In recent years there has been a lot of interests in the connections between
the classical
extended conformal algebras and the Hamiltonian structures of certain classical
integrable systems[1-12]. A protype of connections of
the sort is provided by the
generalized KdV hierarcy which is defined as[2]
$$ {{\partial l_n} \over {\partial t_k}} \equiv \partial_k l_n = [
(l^{k/n}_n)_+, l_n ] \qquad (k=1,2, \dots) \eqno(1.1)$$
where
$$ l_n = \partial^n + u_1 \partial^{n-1}  + u_2 \partial^{n-2} + \dots + u_n
 \eqno(1.2)$$
and  $(A)_{\pm}$ stand for the differential and the integral part of the
pseudodifferential operator $A$ respectively. The second Hamiltonian structure
of (1.1) is given by the second Gelfand-Dickey bracket which, in operator form,
reads
$$\eqalign{ \Theta^{GD}_2 ({{\delta H} \over {\delta l_n}}) &\equiv \{u_1,
H\} \partial^{n-1} + \{u_2, H \} \partial^{n-2} + \dots + \{u_n, H \} \cr
&= (l_n {{\delta H}
\over {\delta l_n}} )_+ l_n  -  l_n ({{\delta H} \over {\delta l_n}} l_n)_+\cr}
\eqno(1.3)$$ Here
$$ {{\delta H} \over {\delta l_n}} = \partial^{-1} {{\delta H} \over {\delta
u_n}} + \partial^{-2} {{\delta H} \over {\delta u_{n-1}}} + \dots +
\partial^{-n} {{\delta H} \over {\delta u_1}}  \eqno(1.4)$$
The connection of this bracket to the classical extended conformal algebra can
be established as follows: First, we define
$$ t(x) = u_2(x) -{{n-1} \over 2} u_1'(x) - {1 \over 2} u_1^2(x) \eqno(1.5)$$
Then it can be shown easily that $t(x)$ satisfies the classical Virasoro
algebra and that $u_1(x)$ is a conformal spin-1 field satisfying the $U(1)$-
$Kac$-$Moody$ algebra[7]; i.e.
$$ \eqalign{ \{t(x), t(y) \} &= [{{n^3-n} \over 12} \partial_x^3 + 2 t(x)
\partial_x + t'(x)] \delta(x-y) \cr
\{u_1(x), t(y) \} &= [ u_1(x) \partial_x + u'_1(x) ] \delta(x-y) \cr
\{u_1(x), u_1(y) \} &= -n \partial_x \delta(x-y) \cr}
\eqno(1.6)$$
Secondly, one can show[10] that for $k(=3, \dots,n)$ a spin-k field $w_k$ can
be
constructed as a differential polynomial of the coefficient functions $u_1,
\dots, u_n$ [We shall review this construction in Sec. II].  Therefore, in
terms of these new fields the second Gelfand-Dickey bracket defines a classical
extended conformal algebra called $W_n$-$U(1)$-$Kac$-$Moody$ algebra. One
usually eliminates the spin-1 field from the spectrum and the resulting algebra
is
the classical $W_n$ algebra. Many generalizations of this connection can be
found in the literature.

Recently there are several publications concerning the so-called ``constrained
KP hierarchy'' from various points of view[13-24]. The constrained hierarchy is
the
ordinary KP hierarchy restricted to pseudodifferential operators of the form
$$ L_n = \partial^n + u_2 \partial^{n-2} + \dots + u_n + \phi
\partial^{-1} \psi \eqno(1.7)$$
The evolution of the system is given by
$$\eqalign{ \partial_k L_n &= [ (L_n^{k/n})_+, L_n ] \cr
            \partial_k \phi &= (L_n^{k/n} \phi)_0, \qquad \partial_k \psi =
 - ((L_n^{k/n})^* \psi)_0  \cr} \eqno(1.8)$$
where $(\quad)_0$ denotes the zeroth order term and $*$ stands for the
conjugate operation: $(AB)^* = B^* A^*, \partial^* = - \partial, f(x)^*=f(x)$.
One should note that the second line of (1.8) is consistent with the first
line. This system can be put into a different Lax representation by using the
gauge transformation
$$\eqalign{ K_n &\equiv \phi^{-1} L_n \phi \cr
  &\equiv \partial^n + v_1 \partial^{n-1} + \dots + v_n + \partial^{-1} v_{n+1}
\cr} \eqno(1.9)$$
In terms of $K_n$ (1.8) becomes the Kuperschmidt's nonstandard Lax hierarchy
[13] (it is also called modified constrained KP hierarchy [18]):
$$ \partial_k K_n = [ (K_n^{k/n})_{\geq 1}, K_n ] \eqno(1.10)$$
The bihamiltonian structures of (1.8) and (1.10) have been constructed by
Oevel and Strampp[18]. The second Hamiltonian structures are somewhat different
from the second
Gelfand-Dickey bracket defined by a pure differential operator. It is a natural
question to examine whether or not they still have the nice conformal property
owned by the bracket defined by (1.2) and (1.3). If yes, what are the
corresponding
conformal algebras? In this paper, we study these two questions in some
details. After a brief review on the conformal property of the second Gelfand-
Dickey bracket defined by a pure differential operator in Sec.II, we show in
Sec.III. that the second Hamiltonian structure defined by (1.7) can be easily
put into a conformally covariant form and gives a nonlocal extended conformal
algebra. We discuss in details two of such nonlocal algebras. We find that
they are very similar to the
V algebra discovered by Bilal[25].
In Sec. IV. we consider the second Hamiltonian
structure defined by (1.9). After working out two examples explicitly, we show
that in terms of the differential operator $l_{n+1} \equiv \partial K_n$ the
second structure is noting but the sum of the second and the third
Gelfand-Dickey brackets defined by $l_{n+1}$. Based on this result we show how
to write $K_n$ in a conformally covariant form. Some concluding remarks are
presented in Sec. V.

\vskip 1cm

\noindent{\bf II. Conformally Covariant Differential Operators}

In this section we review briefly the method to construct the conformally
covariant form of a differential operator[10,11], which will be useful for
later
discussions. First, let us recall a few basic definitions. A function $f(x)$ is
called is spin-k field if it transform under coordinate change $x \rightarrow
t(x)$ as $f(t)= ({dx \over dt})^k f(x)$. The space of all spin-k fields is
denoted by $F_k$. An operator $\Delta$ is called a covariant operator if it
maps from $F_h$ to $F_l$ for some $h$ and $l$. Symbollically, we denote
$ \Delta : F_h \rightarrow F_l$. In other words, under $x \rightarrow t(x)$
$$ \Delta(t) = ({dx \over dt})^l \Delta(x) ({dx \over dt})^{-h}  \eqno(2.1)$$
Usually, $\Delta$ is a pseudodifferential  operator and we can easily write
down the infinitesimal form of (2.1). Taking $t(x)=x - \epsilon(x)$ we can
derive easily the infinitesimal change of $\Delta$[11]:
$$ \delta_{\epsilon} \Delta(x) = \left[ \epsilon(x) \partial_x + l \epsilon'(x)
\right] \Delta(x) - \Delta(x) \left[ \epsilon(x) \partial_x + h \epsilon'(x)
\right] \eqno(2.2)$$

The key step to construct the conformally covariant differential operators is
to recognize that the flow (defined by the second Gelfand-Dickey bracket)
generated by the Virasoro generator $t(x)$ given by (1.5) takes the form of
(2.2). That is, for a suitable choice of $h$ and $l$ a differential operator
$l_n$ can be regarded as a covariant operator mapping from $F_h$ to $F_l$.
It is not hard to work out the values of $h$ and $l$. First, since $l_n$ is
of order $n$, $l-h$ must be $n$. Secondly, the function $u_1$ transforms like
a spin-1 field under the ${\it Virasoro}$ ${\it flow}$
(flow generated by $\int dx
t(x) \epsilon(x)$). One can show easily that $h=-{{n-1} \over 2}$. In short,
we have
$$ l_n \quad : \quad F_{-{{n-1} \over 2}} \longrightarrow F_{{n+1} \over 2}
\eqno(2.3)$$
It is also easy to derive the infinitesimal form of (2.3) by using (1.3),
(1.5), (2.1) and (2.3).

The next step is to construct a family of covariant operators such that
each of them depends on a  spin field and the Virasoro generator $t$.
To this end, we introduce a ``anomalous'' spin-1 field $b(x)$ which obeys the
transformation law: $b(t) = ({dx \over dt}) b(x) + {{d^2 x} \over {dt^2}}
({dx \over dt})^{-1}$. The Virasoro generator is then represented by
$$ t(x) = {{n^3-n} \over 12} \left(b'(x) - {1 \over 2} b(x)^2 \right)
\eqno(2.4)$$
It  is a simpe matter to check that the $t$ represented by (2.4) has the
correct transformation law. The main use of $b$ is to define a sequence of
covariant operators:
$$ D_k^l \equiv [\partial_x -(k+l-1)b(x)][\partial_x - (k+l-2)b(x)] \dots
[\partial_x - k b(x)] \quad (l \geq 0) \eqno(2.5)$$
One may verify that $D_k^l$ maps from $F_k$ to $F_{k+l}$. Now given a
spin-k(=1,2 \dots) field $w_k$ the covariant operator[10]
$$ \Delta_k(w_k,t) \equiv \sum_{l=0}^{n-k} \alpha^{(n)}_{k,l} [D_k^l w_k]
D^{n-k-l} _{-{{n-1} \over 2}} \eqno(2.6)$$
where
$$ \alpha^{(n)}_{k,l} = {{\pmatrix{k+l-1 \cr l} \pmatrix{n-k \cr l}} \over
\pmatrix{2k+l-1 \cr l}} \eqno(2.7)$$
maps from $F_{-{{n-1} \over 2}}$ to $F_{{n+1} \over 2}$. The numerical
coefficients $\alpha^{(n)}_{k,l}$'s are determined by requiring the right hand
side
of (2.6) depends on $b$ only through $t$ defined by (2.4).
With (1.5) and $\alpha^{(n)}_{1,1}={{n-1} \over 2}$ in mind we now can write
down the covariant form
of $l_n$ as $$ l_n = D^{n}_{-{{n-1} \over 2}} + \Delta_1(u_1,t) + {1 \over 2}
\Delta_2(u_1^2,t) + \sum_{k=3}^n \Delta_k(w_k,t) \eqno(2.8)$$
Eq.(2.8) decomposes the coefficient functions $u_k$'s into spin fields and
the Virasoro generator. Inverting these relations then gives the expressions
for
spin fields as differential polynomials of $u_k$'s. The second Gelfand-Dickey
bracket (1.3), when expressed in terms of $t$ and the spin fields $u_1, w_3,
\dots w_n$, defines the $W_n$-$U(1)$-$Kac$-$Moody$ algebra mentioned in
Introduction.

Since the evolution of $u_1$ determined by the Lax equation (1.1) is trivial,
it is often to set $u_1=0$. Under this constraint (2.3) becomes even more
natural
in the sense that it is the only choice which preserves this constraint.
As usual, imposing a constraint causes a modification
of its Hamiltonian structure. For (1.3) it is easy to show following the Dirac
procedure that the modified bracket reads
$$ \bar{\Theta}^{GD}_2({{\delta H} \over {\delta l_n}}) = (l_n {{\delta H}
\over { \delta l_n}})_+ l_n - l_n ({{\delta H} \over {\delta l_n}} l_n)_+
+ {1 \over n} [l_n, \int^x (res[l_n, {{\delta H} \over {\delta l_n}}])]
\eqno(2.9)$$
Here, $u_1$ in $l_n$ and ${{\delta H} \over {\delta u_1}}$ in ${{\delta H}
\over {\delta l_n}}$ are both set to zero and $res(\sum_i a_i \partial^i)
\equiv
a_{-1}$. Now the Virasoro generator $t$ is
simply $u_2$ and the covariant form is still given by (2.8) except that the
$u_1$ dependent terms must be removed. The algebra defined by (2.9) is
the $W_n$ algebra.

Before ending this section we like to remark that the decomposition of a
coefficient function into spin fields given by (2.8) is by no means unique.
Redefinitions like $w_3 \rightarrow w_3 + u_1^3$ and $w_4
\rightarrow w_4 + w_3 u_1$ are certainly allowed. In other words, the
definition of a higher spin field is not unique.
\vfil\eject

\noindent{\bf III. Nonlocal Extended Conformal Algebras From Constrained KP
Hierarchy}

We now consider the Lax operator $L_n$, given by (1.7), for the constrained KP
hierarchy. Here we have set $u_1=0$. For $u_1 \ne 0$ a bihamiltonian
structure associated with $L_n$
has been worked  out by Oevel and Strampp[18]. Using their result and
the Dirac procedure, we can easily write down the second Hamiltonian structure
defined by $L_n$:
$$\eqalign{  \Theta^{KP}_2 ({{\delta H} \over {\delta L_n}}, {{\delta H} \over
{\delta \phi}}, {{\delta H} \over {\delta \psi}})                  &=
(L_n {{\delta H} \over {\delta L_n}} )_+ L_n - L_n ( {{\delta H} \over {\delta
L_n}} L_n )_+ + {1 \over n} [ L_n, \int^x ( res[ L_n, {{\delta H} \over {\delta
L_n}}]) ] \cr
&- \phi \partial^{-1} {{\delta H} \over {\delta \phi}} L_n + L_n {{\delta H}
\over {\delta \psi}} \partial^{-1} \psi + {1 \over n} [ L_n, \int^x (\phi
{{\delta H} \over {\delta \phi}} - \psi {{\delta H} \over {\delta \psi}} ) ]
\cr} \eqno(3.1)$$
Here
$$ {{\delta H} \over {\delta L_n}} =  \partial^{-1} {{\delta H} \over
{\delta u_n}} + \dots + \partial^{-n+1} {{\delta H} \over {\delta u_2}}
\eqno(3.2)$$
The bracket (3.1) is obviously nonlocal. In fact,
$$\eqalign{ \{ \phi(x), \phi(y) \} &= - {{n+1} \over n} \phi(x) \epsilon(x-y)
\phi(y) \cr
\{ \psi(x), \psi(y) \} &= -{{n+1} \over n} \psi(x) \epsilon(x-y) \psi(y) \cr
\{ \phi(x), \psi(y) \} &= {{n+1} \over n} \phi(x) \epsilon(x-y) \psi(y) +
(L_n)_+ \delta(x-y) \cr}
\eqno(3.3)$$
where $\epsilon(x-y) \equiv \partial_x^{-1} \delta(x-y)$ is the antisymmetric
step function.

For $n \geq 2$ it is not hard to check that $t \equiv u_2$ still satisfies the
first of
(1.6). Therefore, (3.1) defines an extended conformal algebra if the higher
spin fields can be constructed. To see this is true, let us try to interpret
$L_n$ as a covariant operator as we did for $l_n$ in Section II.
Since the positive part and the negative part of $L_n$ transform independently
under changes of coordinate,
the constraint $u_1=0$ forces the positive part $(L_n)_+$ maps from $F_{-{{n-1}
\over 2}}$ to $F_{{n+1} \over 2}$. As a result, its negative part does the
same. Hence, we have
$$ L_n \quad : \quad F_{-{{n-1} \over 2}} \longrightarrow F_{{n+1} \over 2}
\eqno(3.4)$$
and, in particular,
$$ \phi(t) \partial_t^{-1} \psi(t) = ({dx \over dt})^{{n+1} \over 2}
\phi(x) \partial_x^{-1} \psi(x) ({dx \over dt})^{{n-1} \over 2} \eqno(3.5)$$
Using $\partial^{-1}_t = \partial_x^{-1} ({dx \over dt})^{-1}$ we obtain
immediately
$$ \phi(t) = ({dx \over dt})^{{n+1} \over 2} \phi(x) \qquad
  \psi(t) = ({dx \over dt})^{{n+1} \over 2} \psi(x) \eqno(3.6)$$
that is, $\phi$ and $\psi$ are both spin-${{n+1} \over 2}$ fields. Thus the
spectrum contains two half integer spin fields if the order $n$  is even.
To complet the discussion we must show that the flow, defined by (3.1),
generated by the functional $\int dx u_2(x) \epsilon(x)$ gives the
infinitesimal form of (3.4).
The verification is completely identical to that for the pure differential
operator
$l_n$ with $u_1=0$ and hence we shall not spell out. From the above discussions
it should be clear that the covariant form of $L_n$ is given by
$$ L_n = D^n_{-{{n-1} \over 2}} + \sum_{k=3}^n \Delta_k(w_k,t) + v^+_{{n+1}
\over 2} \partial^{-1} v^-_{{n+1} \over 2} \eqno(3.7)$$
where $v^{\pm}_{{n+1} \over 2}$ denote two fields of spin-${{n+1} \over 2}$.

Let us now consider two simplest examples. Using (3.7) with $n=2$ we have
$$ L_2 = \partial^2 + t + v^+_{3 \over 2} \partial^{-1} v^-_{3 \over 2}
\eqno(3.8)$$
In this parametrization (3.1) gives[18]
$$ \eqalign{ \{ t(x), t(y) \} &= [{1 \over 2} \partial^3_x + 2 t(x) \partial_x
+ t'(x) ] \delta(x-y) \cr
\{ v^{\pm}_{3 \over 2} (x), t(y) \} &= [{3 \over 2} v^{\pm}_{3 \over 2} (x)
\partial_x + {v^{\pm}_{3 \over 2}}' (x) ] \delta(x-y) \cr
\{ v^{\pm}_{3 \over 2}(x), v^{\pm}_{3 \over 2}(y) \} &= - {3 \over 2} v^{\pm}_
{3 \over 2}(x) \epsilon(x-y) v^{\pm}_{3 \over 2}(y) \cr
\{ v^{\pm}_{3 \over 2}(x), v^{\mp}_{3 \over 2}(y) \} &= {3 \over 2} v^{\pm}_
{3 \over 2}(x) \epsilon(x-y) v^{\mp}_{3 \over 2}(y) \pm [\partial_x^2
+ t(x) ] \delta(x-y) \cr} \eqno(3.9)$$
This is a nonlocal extension of Virasoro algebra by two spin-${3 \over 2}$
fields. (3.9)  is quite similar to Bilal's V algebra which is a nonlocal
extension of Virasoro algebra by two spin-2 fields $v^{\pm}_2$[25]:
$$\eqalign{ \{t(x), t(y) \} &= [ {1 \over 2} \partial^3_x + 2 t(x) \partial_x
+ t'(x)] \delta(x-y) \cr
\{v_2^{\pm}(x), t(y) \} &= [2 v_2^{\pm}(x) + {v_2^{\pm}}'(x) ] \delta(x-y) \cr
\{v_2^{\pm}(x), v_2^{\pm}(y) \} &= -2 v_2^{\pm}(x) \epsilon(x-y) v_2^{\pm}(y)
\cr
\{v_2^{\pm}(x), v_2^{\mp}(y) \} &= 2 v_2^{\pm}(x) \epsilon(x-y) v_2^{\mp}(y)
+ [{1 \over 2} \partial^3_x + 2 t(x) \partial_x + t'(x)] \delta(x-y) \cr}
\eqno(3.10)$$
Next, we consider $n=3$. In this case, the covariant Lax operator reads
$$ L_3 = \partial^3 + t \partial + w_3 + {1 \over 2} t' + v_2^+
\partial^{-1} v_2^- \eqno(3.11)$$
and we have
$$\eqalign{ \{t(x), t(y) \} &=[ 2 \partial^3_x + 2 t(x) \partial_x + t'(x)]
\delta(x-y) \cr
\{w_3(x), t(y) \} &= [3 w_3(x) \partial_x + w'_3(x) ] \delta(x-y) \cr
\{v_2^{\pm}(x), t(y) \} &= [2 v_2^{\pm}(x) + {v_2^{\pm}}'(x) ] \delta(x-y) \cr
\{ w_3(x), w_3(y) \} &= -[{1 \over 6} \partial^5_x + {5 \over 6}
t(x) \partial_x^3 + {5 \over 4} t'(x) \partial_x^2 + ({3 \over 4} t''(x) +
{2 \over 3} t(x)^2 ) \partial_x +
{1 \over 6} t'''(x) \cr
&+ {2 \over 3} t(x)t'(x)
 ] \delta(x-y) + [ 4 v_2^+(x) v_2^-(x) \partial_x + 2 (v_2^+(x) v_2^-(x))']
\delta(x-y) \cr
\{v_2^{\pm}(x), w_3(y) \} &= \pm [{5 \over 3} v_2^{\pm} \partial_x^2  + {5
\over 2} {v_2^{\pm}}'(x) \partial_x + {v_2^{\pm}}''(x) + {2 \over 3} t(x)
v_2^{\pm}(x) ] \delta(x-y) \cr
\{v_2^{\pm}(x), v_2^{\pm}(y) \} &= - {4 \over 3} v_2^{\pm}(x) \epsilon(x-y)
v_2^{\pm}(y) \cr
\{v_2^{\pm}(x), v_2^{\mp}(y) \} &= {4 \over 3} v_2^{\pm}(x) \epsilon(x-y)
v_2^{\mp}(y) + [\partial_x^3 + t(x) \partial_x + {1 \over 2} t'(x) \pm w_3(x)]
\delta(x-y) \cr} \eqno(3.12)$$
Eqs.(3.12) gives some sort of spin-3 extension of V algebra (3.11).
However, one
should note that the elimination of the spin-3 field $w_3$ from (3.12) is not
quite well-defined even though the eliminations of $v_2^{\pm}$ is quite
trivial. It is perhaps better to regard (3.12) as a nonlocal extension of
$W_3$ algebra by two spin-2 fields $v_2^{\pm}$.

We have seen that the bracket (3.1) indeed defines a family of nonlocal
extended algebra when $n \geq 2$. Let us go back to the $n=1$ case:
$$ L_1 = \partial + \phi \partial^{-1} \psi \eqno(3.13)$$
{}From (3.3) we have the following:
$$ \eqalign{ \{ \phi(x), \phi(y) \} &= -2 \phi(x) \epsilon(x-y) \phi(y) \cr
 \{ \psi(x), \psi(y) \} &= -2 \psi(x) \epsilon(x-y) \psi(y) \cr
\{ \phi(x), \psi(y) \}&= \partial_x \delta(x-y) + 2 \phi(x) \epsilon(x-y)
\psi(y) \cr} \eqno(3.14)$$
Here we do not have a natural candidate for the Virasoro generator $t$.
Based on the dimension consideration one  expects that a function
of the form $a \phi^2 + b \phi \psi + c \psi^2 + e \phi' + f \psi'$ could
do the job.  Since $\{t(x), t(y) \}$ contains only local terms, a
little thinking tells us that $a,c,e$ and $f$ must be all zero. Moreover,
simple calculations give
$$\eqalign{ \{\phi(x) \psi(x), \phi(y) \psi(y) \} &= [ 2 \phi(x) \psi(x)
\partial_x + (\phi(x) \psi(x) )' ] \delta(x-y) \cr
\{ \phi(x), \phi(y) \psi(y) \} &= [ \phi(x) \partial_x + \phi'(x) ]
\delta(x-y)\cr
\{ \psi(x), \phi(y) \psi(y) \} &= [ \psi(x) \partial_x + \psi'(x)] \delta(x-y)
\cr} \eqno(3.15)$$
Hence, $\phi$ and $\psi$ are spin-1 fields with respect to the Virasoro
generator which is simply their product $\phi \psi$. This is fairly interesting
situation. We shall come back to this example.
\vskip 1cm

\noindent{\bf IV. $W_n$ Algebras From Nonstandard Lax Operators}

We now consider the Kuperschmidt's nonstandard Lax hierarchy defined by (1.9)
and (1.10). The second Hamiltonian structure for this system is[18]
$$\eqalign{ \Theta_2^{NS} ({{\delta H} \over {\delta K_n}}) &= (K_n {{\delta H}
\over {\delta K_n}})_+ K_n - K_n ( {{\delta H} \over {\delta K_n}} K_n)_+
+ [ K_n, (K_n {{\delta H} \over {\delta K_n}})_0 ] \cr
&+ \partial^{-1} res[K_n, {{\delta H} \over {\delta K_n}}] K_n +
[K_n, \int^x (res[ K_n, {{\delta H} \over {\delta K_n}} ) ] \cr} \eqno(4.1)$$
Here
$$ {{\delta H} \over {\delta K_n}} = {{\delta H} \over {\delta v_{n+1}}} +
\partial^{-1} {{\delta H} \over {\delta v_n}} + \dots + \partial^{-n}
{{\delta H} \over {\delta v_1}} \eqno(4.2)$$
As  in Section III. we like to understand the conformal property of $K_n$ and
the algebra defined by the bracket (4.1). In view of the gauge transformation
$$ K_n = \phi^{-1} L_n \phi = \partial^n + n {\phi' \over \phi} \partial^{n-1}
+ \left(u_2 + {{n(n-1)} \over 2} {\phi'' \over \phi} \right)   \partial^{n-2} +
\dots \eqno(4.3)$$
and of (3.4) and (3.6) we dedue that
$$   K_n \quad : \quad F_{-n} \longrightarrow F_0 \eqno(4.4)$$
and that
$$t \equiv u_2 = v_2 - {{n-1} \over 2} v_1' - {{n-1} \over 2n} v_1^2
\eqno(4.5)$$
is the corresponding Virasoro generator. However, (4.4) and (4.5) are not what
we are interested in since with respect to this $t$ the first coefficient
function $v_1$ is not a spin-1 field. [{\it Recall:} It is a spin-1 field only
when $K_n$ maps from $F_{-{{n-1} \over 2}}$ to $F_{{n+1} \over 2}$. ]
The last statement is actually quite obvious since $v_1 = n {\phi'
\over \phi}$ which is surely not a spin field with respect to $t=u_2$.
As a result (4.1) does not define an extended conforaml algebra with the
choice of Virasoro generator given by (4.5) even though  it has a Virasoro
subalgebra. So if we want to know whether (4.1) defines an extended conformal
algebra or not then we have to give up (4.4), which is the only way $K_n$ can
be a covariant operator, and search for a new Virasoro generator of the form:
$t= v_2 + a v_1' + b v_1^2$. Let us start with two explicit examples.
First, we look at
$$ K_1 = \partial + v_1 + \partial^{-1} v_2 \eqno(4.6)$$
The second Hamiltonian structure associated with $K_1$ has been discussed
in details in Ref.[23]. Here we have
$$\eqalign{ \{v_1(x) ,v_1(y) \} &= 2 \partial_x \delta(x-y) \cr
\{v_1(x), v_2(y) \} &= [\partial^2_x + v_1(x) \partial_x + v_1'(x)] \delta(x-y)
\cr \{v_2(x), v_2(y) \} &= [2 v_2(x) \partial_x + v'_2(x)] \delta(x-y) \cr}
\eqno(4.7)$$
We see that $v_2$ satisfies the Virasoro algebra with zero ``anomalous term'',
$\partial_x^3 \delta(x-y)$. If we take $v_2$ as the Virasoro algebra then $v_1$
is not
a spin-1 field due to the presence of an anomalous term, $\partial^2_x
\delta(x-y)$ in the bracket $\{v_1(x), v_2(y) \}$. However, it was observed
that[23] if we define $t=v_2 + {1 \over 2} v_1'$ then
$$\eqalign{ \{v_1(x), t(y) \} &= [ v_1(x) \partial_x + v_1'(x)] \delta(x-y) \cr
\{ t(x), t(y) \} &= [{1 \over 2} \partial_x^3 + 2 t(x) \partial_x + t'(x)]
\delta(x-y) \cr } \eqno(4.8)$$
In other words, with the new choice of Virasoro generator $v_1$ become a
genuine spin-1 field and the second Hamiltonian structure associated with
$K_1$ is nothing but the $Virasoro$-$U(1)$-$Kac$-$Moody$ algebra.

Next, we consider
$$ K_2 = \partial^2 + v_1 \partial + v_2 + \partial^{-1} v_3 \eqno(4.9)$$
The brackets needed for our discussions are
$$\eqalign{ \{v_1(x), v_1(y) \} &= 6 \partial_x \delta(x-y) \cr
\{v_2(x), v_1(y) \} &= 4 v_1(x) \partial_x \delta(x-y) \cr
\{v_2(x), v_2(y) \} &= [2 \partial_x^3 + 2 (v_2(x) + v_1(x)^2) \partial_x
+ (v_2(x) + v_1(x)^2)' ] \delta(x-y) \cr
\{v_3(x), v_1(y) \} &= [2 \partial_x^3 -2 v_1(x) \partial_x^2 + 2(v_2(x)
 - v_1'(x) ) \partial_x ] \delta(x-y) \cr
\{v_3(x), v_2(y) \} &= [- \partial_x^4 + \partial_x^3 v_(x) + \partial_x v_1(x)
\partial_x^2 - \partial_x v_1(x) \partial_x v_1(x) \cr
&\quad - v_2(x) \partial_x^2 + v_2(x) \partial_x v_1(x) + 3 v_3(x) \partial_x
+ v_3'(x) ] \delta (x-y) \cr} \eqno(4.10)$$
Some straightforward algebras show that if we define
$$ t \equiv v_2 - {1 \over 4} v_1^2   \qquad w_3 \equiv v_3 + {1 \over 2} v_2'
- {1 \over 3} v_1'' - {1 \over 3} v_1 v_2 + {1 \over 12} v_1^3 \eqno(4.11)$$
then
$$\eqalign{ \{v_1(x), t(y) \} &= [v_1(x) \partial_x + v_1'(x) ] \delta(x-y) \cr
\{t(x), t(y) \} &= [2 \partial_x^3 + 2 t(x) \partial_x + t'(x) ] \delta(x-y)
\cr
\{w_3(x), t(y) \} &= [3 w_3(x) \partial_x + w_3'(x) ] \delta(x-y) \cr}
\eqno(4.12)$$
That is, $v_1$ and $w_3$ are, respectively, spin-1 and spin-3 fields with
respect to the Virasoro generator. In terms of $v_1$, $t$ and $w_3$ we expect
the second Hamiltonian structure associated with $K_2$ defines the $W_3$-$U(1)$
-$Kac$-$Moody$ algebra. These two examples suggest that for any $K_n$ a new
Virasoro generator and a new set of higher spin fields can be found such that
(4.1) gives the $W_{n+1}$-$U(1)$-$Kac$-$Moody$ algebra.

Let us now show that the above expectation is indeed true. The key hint comes
from an observation on the coefficient of the anamolous term $\partial_x^3
\delta(x-y)$ in the bracket $\{ t(x), t(y) \}$. From (4.8) and (4.12) we see
that it  is ${1 \over 2}$ for $K_1$ and is 2 for $K_2$. These two values are
precisely given by the formula ${{n^3-n} \over 12}$ [see (1.6)] with $n=2$ and
$n=3$ respectively. It is then natural to suspect that the conformal algebras
associated with $K_1$ and $K_2$ might be related to the conformal property
of the
pure differential operators $l_2$ and $l_3$. We are therefore motivated to
consider
the following mapping:
$$\eqalign{ l_{n+1} \equiv  \partial K_n &= \partial^{n+1} + v_1 \partial^n
+ (v_2 + v_1') \partial^{n-1} + \dots    + (v_{n+1} + v_n') \cr
&\equiv \partial^{n+1} + u_1 \partial^n + u_2 \partial^{n-1} + \dots + u_{n+1}
\cr}
\eqno(4.13)$$
{}From (4.13) it can be proved rigorously[see Appendix] that effectively
$$ \eqalign{ {{\delta H} \over {\delta l_{n+1}}} &\equiv \partial^{-1}
{{\delta H} \over {\delta u_{n+1}}} + \dots + \partial^{-n-2} {{\delta H} \over
{\delta u_1}} \cr
&= {{\delta H} \over {\delta K_n}} \partial^{-1} \cr} \eqno(4.14)$$
A simple way to understand (4.14) is to note that it implies $ tr(l_{n+1}
{{\delta H} \over {\delta l_{n+1}}} ) = tr (K_n {{\delta H} \over {\delta K_n}}
)$. [{\it Recall:} $tr(A) \equiv \int dx {\rm } res(A)$ ] Now we like to
express
(4.1) in terms of $l_{n+1}$ and ${{\delta H} \over
{\delta l_{n+1}}}$. By the use of (4.13) and (4.14) we easily derive
$$ \eqalign{ (l_{n+1} {{\delta H} \over {\delta l_{n+1}}} )_+ &=
(K_n {{\delta H} \over {\delta K_n}} )_+  + [ \partial, (K_n {{\delta H} \over
{\delta K_n}} )_{\geq 1} ] \partial^{-1} \cr
(l_{n+1} {{\delta H} \over {\delta l_{n+1}}} )_+ l_{n+1} &= \partial (K_n
{{\delta
H} \over {\delta K_n}} )_+ K_n - (K_n {{\delta H} \over {\delta K_n}} )'_0 K_n
\cr res[ l_{n+1}, {{\delta H} \over {\delta l_{n+1}}} ] &=
res[ K_n, {{\delta H} \over {\delta K_n}} ] + ( K_n {{\delta H} \over {\delta
K_n}} )'_0 \cr} \eqno(4.15)$$
Eqs.(4.15) then lead to
$$ \eqalign{ \partial \Theta_2^{NS} ({{\delta H} \over {\delta K_n}}) &=
(l_{n+1} {{\delta H} \over {\delta l_{n+1}}} )_+ l_{n+1} - l_{n+1}
({{\delta H} \over {\delta l_{n+1}}} l_{n+1} )_+ + [ l_{n+1}, \int^x (res[
l_{n+1}, {{\delta H} \over {\delta l_{n+1}}} ] ) ] \cr
&\equiv \wp ( {{\delta H} \over {\delta l_{n+1}}} ) \cr
} \eqno(4.16)$$
The last piece in the first line of (4.16) is called the third Gelfand-Dickey
bracket[10]. We have
shown that under the mapping (4.13) the Hamiltonian structure (4.1) is
transformed to the sum of the second
and the third Gelfand-Dickey brackets defined by the pure differential operator
$l_{n+1}$ of order $n+1$.

With (4.16) the remained tasks are quite easy. We like to find a Virasoro
generator, which is a differential polynomial in the coefficients of $l_n$,
such that the
corresponding flow defined by (4.16) gives the infinitesimal form of (3.4);i.e.
$$ \delta_{\epsilon} l_n = [ \epsilon \partial + {{n+1} \over 2} \epsilon']
l_n  - l_n [ \epsilon \partial - {{n-1} \over 2} \epsilon'] \eqno(4.17)$$
We find that
$$ t = u_2 - {{n-1} \over 2} u_1' -{{n-2} \over {2(n-1)}} u_1^2 \eqno(4.18)$$
does the job. $u_1$ and $t$ together satisfy (1.6) except that the last
bracket there is now replaced by
$$ \{u_1(x), u_1(y) \} = n(n-1) \partial_x \delta(x-y) \eqno(4.19)$$
Since $v_1=u_1$ according to (4.13), (4.19) agrees with (4.7) and (4.10).
Eq.(4.18) enables us the write down the covariant form of $l_n$ with respect
to the bracket defined by (4.16):
$$ l_n = D^n_{-{{n-1} \over 2}} + \Delta_1(u_1,t) + {{n-2} \over {2(n-1)}}
\Delta_2(u_1^2,t) + \sum^n_{k=3} \Delta_k(w_k,t) \eqno(4.20)$$
The formula (4.20) differs from (2.8) only by the coefficient of the third
term. Let us compare the decompositions given by (4.20) with the previous
explicit results for $K_1$ and $K_2$ . In terms of $t$ and
spin fields we have
$$ \eqalign{ l_2 \equiv \partial K_1 &= \partial^2 + v_1 \partial + t + {1
\over 2} v_1'
\cr
l_3 \equiv \partial K_2 &= \partial^3 + v_1 \partial^2 + (t + v_1' + {1 \over
4} v_1^2)
\partial \cr
&\qquad + w_3 + {1 \over 2} t' + {1 \over 3} v_1'' + {1 \over 3} v_1 t +
{1 \over 4} v_1 v_1' \cr} \eqno(4.21)$$
It is easy to check that (4.21) completely agrees with (4.20) once $v_1=u_1$ is
considered.

Finally, we like to impose the constraint $u_1=0$ on (4.16). By the virtue of
(4.13) this is essentially equivalent to imposing the constraint $v_1=0$ on
(4.1). We
leave the discussion of this equivalence to the
Appendix. Following the Dirac procedure the modified form of $\wp$
defined by (4.16) reads
$$\bar{\wp} ({{\delta H} \over {\delta l_{n+1}}}) = (l_{n+1} {{\delta H} \over
{\delta
l_{n+1}}} )_+ l_{n+1} -l_{n+1} ({{\delta H} \over {\delta l_{n+1}}} l_{n+1})_+
+ { 1\over {n+1}}
[ l_{n+1}, \int^x ( res[ l_{n+1}, {{\delta H} \over {\delta l_{n+1}}}] ) ]
\eqno(4.22)$$
$\bar{\wp}$ is exactly equal to the modified second Gelfand-Dickey bracket
given
by (2.9). This shows that when we remove the spin-1 field $u_1$ the resulting
algebra is simply the $W_{n+1}$ algebra. This result together with the fact
that
$u_1$ and $t$ together define the $U(1)$-$Kac$-$Moody$ algebra confirm our
expectation that (4.1) [or equivalently (4.16)] defines  the
$W_{n+1}$-$U(1)$-$Kac$ -$Moody$ algebra. However, it is not equivalent to
the corresponding one defined by (1.3) and (2.8) in the sense that no
redefinitions, of differential polynomial type, of the Virasoro generator and
of spin fields makes two algebras identical(We assume that $u_i$'s are real-
valued functions). This inequivalence actually can
be easily seen by comparing the last of (1.6) and (4.19): no redefinitions
of $u_1$
and of $t$ can make two $Virasoro-U(1)$-$Kac$-$Moody$ subalgebras identical.

\vskip 1cm

\noindent{\bf V. Concluding Remarks}

In this paper we have studied the conformal property of the Lax operator of
the constrained KP hierarchy and the associated second Hamiltonian structure.
We have seen that the analysis for
the second Gelfand-Dickey bracket defined by a pure differential operator can
be straightforwardly carried over
to the present case. The conformal decomposition defined by (3.7) is a simple
extension of (2.8). The extended conformal algebras defined by (3.1) are
nonlocal and contain two half integer spin fields when the leading order is
even. We have given two examples of such nonlocal algebras. They are very
similar to Bilal's V algebra.

   We also study the constrained KP
hierarchy in Kuperschmidt's nonstandard Lax representation. Here we have found
that the corresponding second Hamiltonian structure defines the
$W_{n+1}$-$U(1)$-
$Kac$-$Moody$ algebra just like the second Gelfand-Dickey bracket of the
differential operator of order $n+1$ does.  However, the Virasoro generator
in this algebra does not make its Lax operator ,$K_n$, a covariant operator.
The natural covariant operator now is actually the pure differential operator
$l_{n+1} \equiv \partial K_n$ of order $n+1$. More unexpectedly, in terms of
$l_{n+1}$, the Hamiltonian structure (4.1) becomes the sum of the second
and the third Gelfand-Dickey brackets defined by $l_{n+1}$. The conformally
covariant form is then worked out. Since the two brackets (3.1) and (4.1) are
connected by the gauge transformation (1.9), what is interesting here is that a
nonlocal extended conformal algebra defined by (3.1) is nothing but a local
$W$-$U(1)$-$Kac$-$Moody$ algebra in disguise. Of course, the transfomations
between the two sets of Virasoro generator and spin fields are not of
differential polynomial type. For instance, connecting (3.13) to (4.6) by
$K_1 = \phi^{-1} L_1 \phi$ we have
$$ \eqalign{ v_1 &= {{\phi'} \over \phi} \cr
 t &= \phi \psi + {1 \over 2} ( {\phi' \over \phi} )' \cr} \eqno(5.1)$$
which maps the nonlocal algebra (3.14) to the local one given by (4.7) and
(4.8).

\vskip .5cm

\noindent{\bf Acknowledgment}

We thank Professor Ashok Das for valuable comments.
The work was supported by the Natiional Science Council under Grant number
NSC-84-2112-M-007-017.

\vskip 1cm

\noindent{\bf Appendix}

In this appendix we give a proof of (4.14). From (4.13) we have
$$ \eqalign{ v_1 &= u_1 \cr
v_2 &= u_2 - u_1' \cr
v_3 &= u_3 - u_2' + u_1'' \cr
&\dots \dots      \cr
&\dots \dots \dots     \cr
v_{n+1} &= u_{n+1} - u_n' + \dots + (-1)^n u_1^{(n)} \cr} \eqno(A.1)$$
and hence
$$\eqalign{ {{\delta H} \over {\delta u_1}} &= {{\delta H} \over {\delta v_1}}
+ ( {{\delta H} \over {\delta v_2}} )' + \dots + ({{\delta H} \over {\delta
v_{n+1}}} )^{(n)} \cr
&\dots \dots \dots \cr
&\dots \dots \cr
{{\delta H} \over {\delta u_n}} &= {{\delta H} \over {\delta v_n}} +
( {{\delta H} \over {\delta v_{n+1}}} )' \cr
{{\delta H} \over {\delta u_{n+1}}} &= {{\delta H} \over {\delta v_{n+1}}} \cr}
\eqno(A.2)$$
Using
$$ f \partial^{-1} = \partial^{-1} f + \partial^{-2} f' + \partial^{-3} f''
+ \dots \eqno(A.3)$$
we obtain
$$\eqalign{ {{\delta H} \over {\delta K_n}} \partial^{-1} &=
\left( {{\delta H} \over {\delta v_{n+1}}} + \partial^{-1} {{\delta H} \over
{\delta v_n}} + \dots + \partial^{-n} {{\delta H} \over {\delta v_1}} \right)
\partial^{-1} \cr
&= \partial^{-1} {{\delta H} \over {\delta v_{n+1}}} + \partial^{-2} \left[
 {{\delta H} \over {\delta v_n}} + ({{\delta H} \over {\delta v_{n+1}}} )'
 \right] + \dots \cr
&\quad+ \partial^{-n-1} \left[ {{\delta H} \over {\delta v_1}} + ({{\delta H}
\over {\delta v_2}} )' + \dots + ({{\delta H} \over {\delta v_{n+1}}})^{(n)}
 \right] + O( \partial^{-n-2} ) \cr
&= \partial^{-1} {{\delta H} \over {\delta u_{n+1}}} + \partial^{-2} {{\delta
H} \over {\delta u_n}} +  \dots + \partial^{-n-1} {{\delta H} \over {\delta
u_1}} + O(\partial^{-n-2}) \cr
&= {{\delta H} \over {\delta l_{n+1}}} + O(\partial^{-n-2}) \cr} \eqno(A.4)$$
Since terms in  $O(\partial^{-n-2})$ do not contribute to (4.15) or (4.16),
we can simply drop them. This completes the proof for (4.14).

Now let us imposing constraint $v_1=0$ on (4.1). The Dirac procedure now gives
the modified bracket:
$$\eqalign{ \bar{\Theta}^{NS}_2 ({{\delta H} \over {\delta K_n}} ) &=
(K_n {{\delta H} \over {\delta K_n}} )_+ K_n - K_n ({{\delta H} \over {\delta
K_n}} K_n )_+ + {1 \over {n+1}} \big\{ [ K_n, (K_n {{\delta H} \over {\delta
K_n}} )_0 ] \cr
&+ [ K_n, \int^x (res[ K_n, {{\delta H} \over {\delta K_n}} ] ) ]
+ \partial^{-1} res[K_n, {{\delta H} \over {\delta K_n}} ] K_n
- n \partial^{-1} (K_n {{\delta H} \over {\delta K_n}})'_0 K_n \big\} \cr}
\eqno(A.5)$$
As in Section IV. we like to express (A.5) in terms of $l_{n+1}$ and ${{\delta
H} \over {\delta l_{n+1}}}$. Here some care must be taken. From (A.2) we
see that ${{\delta H} \over {\delta v_1}}=0$ does not imply ${{\delta H} \over
{\delta u_1}} = 0$. As a consequence, instead of (A.4) we have
$$ \eqalign{ {{\delta H} \over {\delta K_n}} \partial^{-1} &= \partial^{-1}
{{\delta H} \over {\delta u_{n+1}}} + \dots + \partial^{-n} {{\delta H} \over
{\delta u_2}} + O_{-n-1} + O( \partial^{-n-2} ) \cr
&\equiv {{\delta H} \over {\delta l_{n+1}}} + O(\partial^{-n-1}) \cr }
\eqno(A.6)$$
where $O_{-n-1}$ denotes terms of order $-n-1$. Since the term $O_{-n-1}$
could affect (4.15), in order to use (4.15) the symbol ${{\delta H} \over
{\delta l_{n+1}}}$ there
must be regarded as the sum of ${{\delta H} \over {\delta l_{n+1}}}$ and
$O_{-n-1}$ defined in (A.6). A derivation similar to that for (4.16) then
gives
$$ \partial \bar{\Theta}^{NS}_2 = (l_{n+1} {{\delta H} \over {\delta l_{n+1}}}
)_+ l_{n+1} - l_{n+1} ({{\delta H} \over {\delta l_{n+1}}} l_{n+1} )_+
+ {1 \over {n+1}} [ l_{n+1}, \int^x (res[ l_{n+1}, {{\delta H} \over {\delta
l_{n+1}}} ] ) ] \eqno(A.7)$$
We observe that the term $O_{-n-1}$,
which was previously put into ${{\delta H} \over {\delta l_{n+1}}}$, does
not contribute to (A.7). Hence we can drop it and the equality (4.14) is again
correct. As promised, (A.7) is identical to (4.22). We thus have shown that
imposing the constraint
$v_1=0$ on (4.1) is completely equivalent to imposing the constraint $u_1=0$
 on (4.16).

\vskip 1cm

\noindent{\bf References}

\item{[1]} A. Das, {\it Integrable Models} (World Scientific, 1989).
\item{[2]} L.A. Dickey, {\it Soliton Equations and Hamiltonian Systems} (World
Scientific, 1991).
\item{[3]} A.B. Zamolochikov, Theor. Math. Phys. {\bf 65}, 1205 (1985).
\item{[4]} J.-L. Gervais, Phys. Lett. {\bf B160}, 277 (1985).
\item{[5]} P. Mathieu, Phys. Lett. {\bf B208}, 101 (1988).
\item{[6]} I. Bakas, Nucl. Phys. {\bf B302}, 189 (1988); Phys. Lett. {\bf
B213}, 313 (1988).
\item{[7]} I. Bakas and A.C. Kakas, J. Phys. {\bf A22}, 1451 (1989).
\item{[8]} Q. Wang, P.K. Panigrahi, U. Sukhatme and W.Y. Keung, Nucl. Phys.
{\bf B344}, 196 (1990).
\item{[9]} A. Das and S. Roy, Int. J. Mod. Phys. {\bf A6}, 1429 (1991);
J. Math. Phys. {\bf 32}, 869 (1991).
\item{[10]} P. Di Francesco, C. Itzykson and J.-B. Zuber, Comm. Math. Phys.
{\bf 140}, 543 (1991).
\item{[11]} W.-J. Huang, J. Math. Phys. {\bf 35}, 993 (1994).
\item{[12]} J.M. Figueroa-O'Farrill, J. Mas and E. Ramos, Phys. Lett. {\bf
B299}, 41 (1993) and references therein.
\item{[13]} B.A. Kuperschmidt, Comm. Math. Phys. {\bf 99}, 51 (1985).
\item{[14]} A. Orlov, {\it Symmetries for Unifying Different Soliton Systems
into a Single Hierarchy}, Preprint IINS/Oce-04/03, Moscow (1991).
\item{[15]} Bing Xu and Y. Li, J. Phys. {\bf A25}, 2957 (1992).
\item{[16]} Bing Xu, A Unified Approach to Recursion Operators of the Reduced
1+1 Dimensional Systems, Preprint, Hefei (1992).
\item{[17]} Y. Cheng, J. Math. Phys. {\bf 33}, 3774 (1992).
\item{[18]} W. Oevel and W. Strampp, Comm. Math. Phys. {\bf 157}, 51 (1993).
\item{[19]} L. Dickey, {\it On the Constrained KP Hierarchy}, Oklahoma preprint
(hep-th/9407038).
\item{[20]} H. Aratyn, E. Nissimov and S. Pacheva, Phys. Lett. {\bf B314}, 41
(1993).
\item{[21]} H. Aratyn, L.A. Ferreira, J.F. Gomes and A.H. Zimmerman, Nucl.
Phys. {\bf B402}, 85 (1993); {\it On W(infinity) Algebras, Gauge Equivalence of
KP Hierarchy, Two Boson Realizations and Their KdV Reductions}, preprint
(hep-th/9304152).
\item{[22]} L. Bonora and C.S. Xiong, Phys. Lett. {\bf B285}, 191 (1992);
Int. J. Mod. Phys. {\bf A8}, 2973 (1993).
\item{[23]} J.C. Brunelli, A. Das and W.-J. Huang, Mod. Phys. Lett. {\bf A9},
2147 (1994).
\item{[24]} H. Aratyn, J.F. Gomes and A.H. Zimmerman, {\it Affine Lie Algebraic
Origin of Constrained KP Hierarchy}, Preprint (hep-th/9408104).
\item{[25]} A. Bilal, Nucl. Phys. {\bf B422}, 258 (1994); {\it Multi-Component
KdV Hierarchy, V Algebra and Non Abelian Toda Theory}, Preprint PUPT-1446
(hep-th/9401167).

\end

\end